# Unraveling the Defect Physics of SiC Micropipe Sidewalls by Non-Line-of-Sight Confocal Spectromicroscopy: Amphoteric Giant Traps

Irwan Saleh Kurniawan[#1,2], Russel Cruz Sevilla[#1,2], Ruth Jeane Soebroto[1,2], Hsiu-Ying Huang[1,2], Hsiu-Ming Hsu[1,2], Ji-Lin Shen[1,2], Sheng Hsiung Chang[3], Wen-Chung Li*[1,2,4] Chi-Tsu Yuan*[1,2]

[1]Department of Physics, Chung Yuan Christian University, Taoyuan, Taiwan

[2]Research Center for Semiconductor Materials and Advanced Optics, Chung Yuan Christian University, Taoyuan, Taiwan

[3]Department of Optics and Photonics, National Central University, Taoyuan, Taiwan

[4]WAFER WORKS, Taoyuan, Taiwan

**Abstract**

Micropipes are among the most detrimental defects in SiC wafer and are closely linked to catastrophic device failure. However, the microscopic defect nature of their internal sidewalls and the mechanism of the associated leakage current remain poorly understood, because their high-aspect-ratio geometry severely restricts direct optical probing. Here, we develop a non-line-of-sight confocal multiple-reflection spectromicroscopy technique combined with direct defect photoionization to unravel the defect physics of micropipe sidewalls. We show that these sidewalls host a high density of donor-like and acceptor-like deep-level states, giving rise to ultrabroad emission bands composed of intrinsic donor–acceptor-pair-like (DAP-like) recombination and detrapping-mediated free-to-bound transitions. Unlike conventional defect luminescence, the DAP-like emission remains dominant even at room temperature across all excitation powers. This behavior is attributed to rapid carrier capture by the sidewall defects, as evidenced by fast-rising and nanosecond-scale decay dynamics, along with coupled carrier kinetics. These results suggest that micropipe sidewalls can serve as extended amphoteric giant traps and carrier reservoirs, facilitating leakage current through trap-assisted transport. Our work provides a nondestructive optical approach for directly probing high-aspect-ratio extended defects and offers deep mechanistic insight into their defect physics and leakage mechanisms.

**Key words:**

Micropipe sidewalls, non-line-of-sight confocal spectromicroscopy, defect physics, amphoteric giant traps, leakage current mechanism

## INTRODUCTION

A variety of crystalline defects still exist in 4H-SiC wafer and can seriously degrade device performance and reliability[1,2]. Among them, micropipes are particularly detrimental, as they can induce severe leakage currents and premature breakdown, leading to catastrophic device failure[3,4]. Even a single micropipe intersecting the active device region can cause device failure. Despite substantial efforts to eliminate micropipes, they still occasionally appear and have been especially problematic in large-diameter and thick epitaxial wafers[5,6]. Although the device-killing impact of micropipes is well recognized, the microscopic defect nature and underlying defect physics associated with leakage current mechanism remain largely unexplored, particularly the deep-level states at the internal sidewalls of their hollow cores[7,8].

Micropipes are superscrew dislocations with large Burgers vectors and comprise a hollow core bounded by highly defective inner sidewalls[9,10]. The strong strain fields, structural disorder, and unsaturated bonds associated with these sidewalls are expected to give rise to a high density of energetically and spatially distributed deep-level states. Such states can strongly interact with charge carriers, thereby degrading device performance and reliability[11]. Recently, H. Das et al. systematically investigated the impact of micropipes on device performance throughout the manufacturing chain and reported very high device-killing ratios for both diodes and MOSFETs[12]. They further showed that some micropipes can partially close or dissociate into threading screw dislocations (TSDs), which can exhibit high reverse leakage currents and ultimately lead to device failure.

H. L. Yan et al. investigated the influence of micropipes on the yield and electrical performance of SiC JBS diodes[13]. They reported that most micropipes propagate through the epitaxial layer to the surface and exhibit significant reverse leakage current and premature breakdown. Interestingly, they also identified another class of micropipes, referred to as "weak micropipes," which could not be readily resolved by commercial inspection equipment, possibly because of their small dimensions or limited optical contrast at the surface, and did not induce measurable leakage current under reverse bias. These observations suggest that the electrical activity of micropipes is not determined solely by their geometry-induced electric-field crowding effect, but may also depend on other microscopic factors, such as the density and nature of deep-level defects associated with the sidewalls.

Thanks to substantial advances in SiC crystal growth, the density of micropipes has been markedly reduced. Nevertheless, micropipes and submicron micropipes still persist and remain a major concern, particularly in thick epitaxial layers and large-diameter wafers. Recently, J. W. Yang et al. reported a new type of micropipe in thick homoepitaxial 4H-SiC layers[7]. Unlike conventional micropipes with cylindrical cores, these defects exhibit a truncated hexagonal morphology while retaining the 4H-SiC polytype, with no evidence of polytype transformation. More recently, A. M. A. Lee et al. employed selected-area electron diffraction (SAED) and energy-dispersive X-ray spectroscopy (EDX) to investigate the micropipes in 200 mm 4H-SiC epitaxial wafer, revealing silicon-rich sidewalls with Si:C stoichiometric ratios as high as ~9:1, in sharp contrast to the nearly stochiometric host SiC[6].

Most previous studies of micropipes have focused primarily on their evolution during crystal growth and their impact on device performance[14-16]. Although micropipes are widely recognized as leakage-current "killer" defects, their electrical behavior has often been straightforwardly interpreted mainly in terms of geometric field crowding effect arising from the hollow-core morphology, whereas the contribution of the defective micropipe sidewalls has received far less attention. If this simplified picture were sufficient, then defects with comparable field-crowding geometries, such as closed-core screw dislocations, and other hollow or recessed defects including voids and pits, would be expected to exhibit similar electrical behavior. However, this expectation is inconsistent with experimental observations and cannot adequately explain the large leakage current at low reverse bias or the premature breakdown behavior[17,18].

Despite their potential significance, the microscopic nature and underlying defect physics of micropipe inner sidewalls remain poorly understood, largely because the hollow-core geometry prevents direct access to the inner surfaces by conventional optical characterization. Consequently, a direct, nondestructive optical approach for probing micropipe inner sidewalls is highly desirable but remains technically challenging.

To address these challenges and uncover the defect physics of micropipe sidewalls, we developed a purely optical, non-line-of-sight confocal multiple-reflection defect-PL spectromicroscopy, together with direct defect photoionization. By leveraging internal optical reflections within the micropipe hollow core, our technique enables direct, nondestructive probing of the defects on the inner sidewalls. Remarkably, we observed ultrabroad, inherent DAP-dominated defect emission emerging from the micropipe cores, providing valuable information regarding the underlying electronic structure and recombination dynamics. We found that micropipe sidewalls can act as extended amphoteric giant traps and carrier reservoirs that efficiently capture both electrons and holes, giving rise to dominant donor–acceptor-pair (DAP)-like emission with unique carrier dynamics, including a rapid rise, a nanosecond-scale decay, and a rebuilt plateau. These findings further suggest that micropipe sidewall defects facilitate leakage-current transport through trap-assisted mechanisms.

## RESULTS

### Non-line-of-sight confocal spectromicroscopy for probing micropipe sidewall defects

To achieve direct optical probing of micropipe sidewalls, we designed a non-line-of-sight confocal multiple-reflection spectromicroscopy technique that exploits the point-probing capability of 3D confocal optics to couple light into the micropipe, as schematically illustrated in Fig. 1. When the excitation laser is tightly focused at a free-space position inside the micropipe, the beam diverges within the hollow core and undergoes multiple reflections from the inner sidewalls. This multipath optical process substantially enhances the photon–defect interactions, thereby enabling simultaneous detection of laser backscattering (BS) and defect-related photoluminescence (DPL).

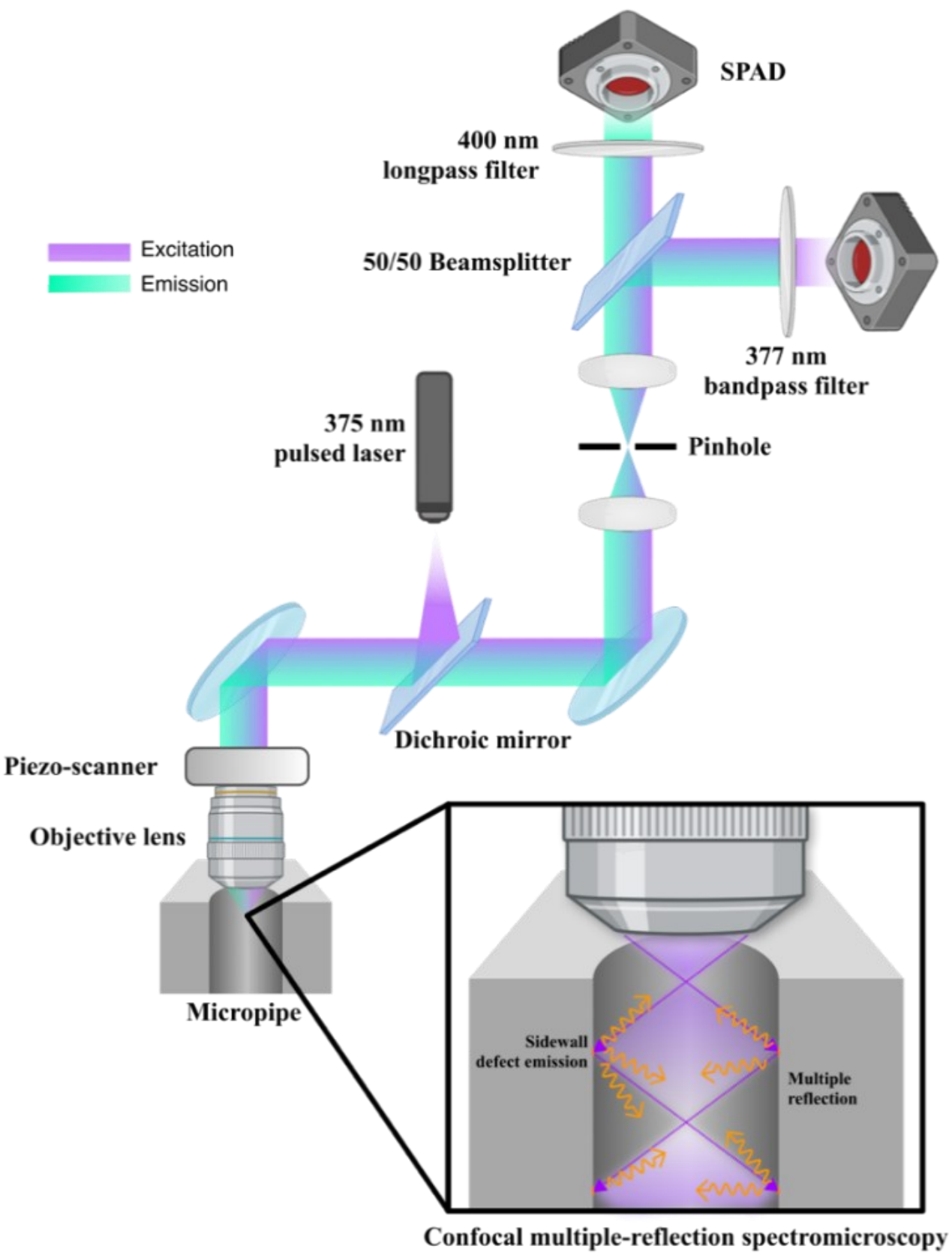

**Fig. 1|** Physical concept and experimental setup of non-line-of-sight confocal multiple-internal-reflection spectromicroscopy.

To demonstrate the capability of our technique, we first performed surface laser BS and DPL mapping of a selected micropipe, as shown in Fig. 2a and 2b. In the surface BS image (Fig. 2a), a bright background is observed, with pronounced dark contrast at the micropipe locations. These dark features arise from reduced backscattering within the hollow cores, thereby confirming the hollow-core nature of the micropipes[19,20]. In the DPL image (Fig. 2b), a long-pass filter (>400 nm) was employed to suppress near-band-edge emission while transmitting defect-related emission. Under these conditions, one would expect minimal contrast against a dark background, since the band-edge emission is largely rejected and empty hollow regions are not expected to produce detectable luminescence.

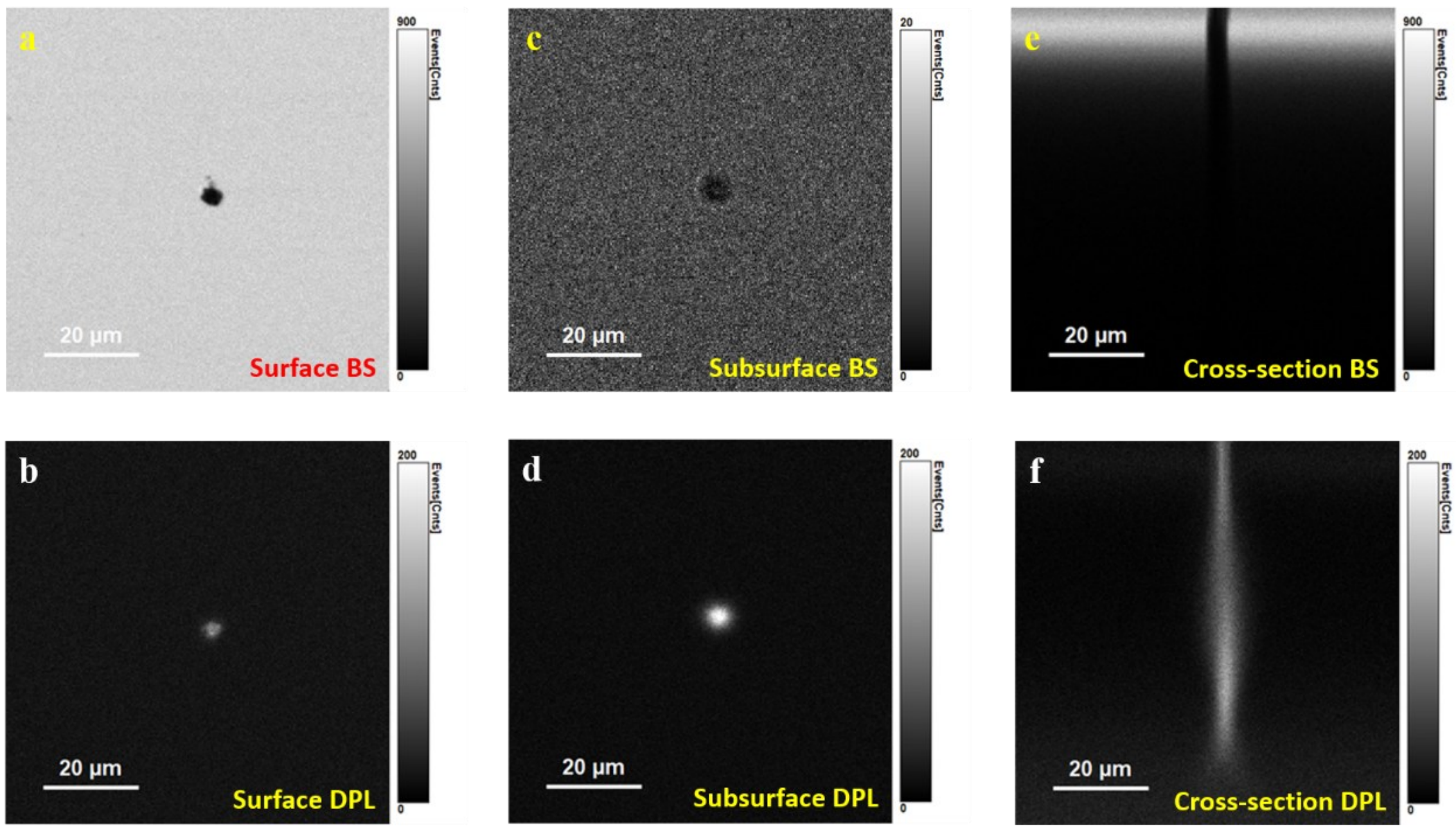

**Fig. 2| Non-line-of-sight confocal dual-mode mapping of micropipes. (a), (b)** Surface BS and DPL mapping images of the micropipes. **(c)–(f)** Subsurface BS/DPL and cross-sectional BS/DPL mappings of the same micropipes.

Surprisingly, distinct bright spots appear at the micropipe locations in the surface DPL image. At first glance, this observation appears counterintuitive, since an empty cavity would not be expected to contain emissive centers under a conventional line-of-sight optical configuration. However, this behavior is fully consistent with the optical response enabled by our non-line-of-sight confocal multiple-reflection-assisted excitation–emission microscopy scheme. The hollow-core geometry supports multiple internal reflections of both the excitation light and the defect-related emission. These repeated reflections efficiently couple the excitation light to the defective inner sidewalls and redirect the resulting DPL into the collection path, thereby enabling direct optical access to inner-sidewall defects that remain challenging to probe using conventional techniques and commercial defect inspection optical systems.

To further verify that the anomalous luminescence originates from the micropipe sidewalls rather than from experimental artifacts, we performed subsurface and cross-sectional BS/DPL mapping on the same micropipes (Fig. 2c–2f). As expected, the subsurface BS image exhibits a pattern like that observed in the surface map, but with reduced contrast owing to suppression of surface specular reflection by the confocal pinhole. In addition, the corresponding DPL map reproduces the hollow-core-associated emission features with enhanced intensity and contrast, confirming that the emission originates from the micropipe sidewalls through non-line-of-sight collection. Notably, the cross-sectional DPL map reveals emission distributed across the hollow-core region, whereas the corresponding BS image displays a pronounced dark column at the micropipe location.

This opposite contrast behavior is fully consistent with the hollow geometry: the empty core contains negligible material and therefore generates minimal volume backscattering, resulting in reduced BS intensity. In contrast, the defect-related emission activated through the multiple-reflection–assisted excitation process can be redirected and partially guided back toward the objective, giving rise to detectable DPL within the hollow-core region. This observation provides robust evidence that the anomalous DPL of the sidewall defects can indeed be detected through the confocal configuration when applied to hollow-core geometries. To further investigate the DPL emission, we performed cross-sectional confocal DPL mapping on the same micropipes using the objectives with different numerical apertures (NAs, 0.4 and 0.65), thereby tuning the excitation and collection geometry through the change of the cone angle of the focused laser beam. In this configuration, the excitation beam undergoes multiple reflections within the hollow core, and the angular distribution of internal propagation is predominantly governed by the objective NA. A lower-NA objective introduces a narrower angular spectrum with a smaller convergence angle, resulting in a longer effective propagation distance along the hollow core. In contrast, a higher-NA objective generates a wider angular cone and stronger beam divergence, leading to a more confined excitation volume and a reduced axial extent of sidewall-coupled interaction.

Fig. 3a and 3b show the lateral DPL mapping images of the micropipe acquired with the objectives of different numerical apertures (NA = 0.4 and 0.65). In both cases, similar contrast behavior was observed, consistent with the hollow-core geometry and sidewall-originated DPL discussed above. As expected, the primary difference between the two lateral mappings is the overall signal level, which decreases with decreased NA due to the reduced collection efficiency. Fig. 3c–3f show cross-sectional BS and DPL images of the same micropipes, acquired by scanning the focal plane from ~5 μm above the surface deep into the bulk. The BS images remain similar for different NA, whereas the DPL images reveal a distinct bright column along the micropipe core, with two notable features. First, a measurable DPL signal is detected even when the nominal focal position is above the surface. Although seemingly counterintuitive, this observation is fully consistent with the multiple-reflection coupling mechanism again: when the beam is focused slightly above the surface, part of the incident light can still couple into the micropipe hollow, propagate along the hollow core, and undergo repeated interactions with the inner sidewalls, thereby exciting and collecting the DPL [21]. Second, as expected, the axial extent of the bright column increases as the NA decreases. Altogether, these observations support our proposed scenario.

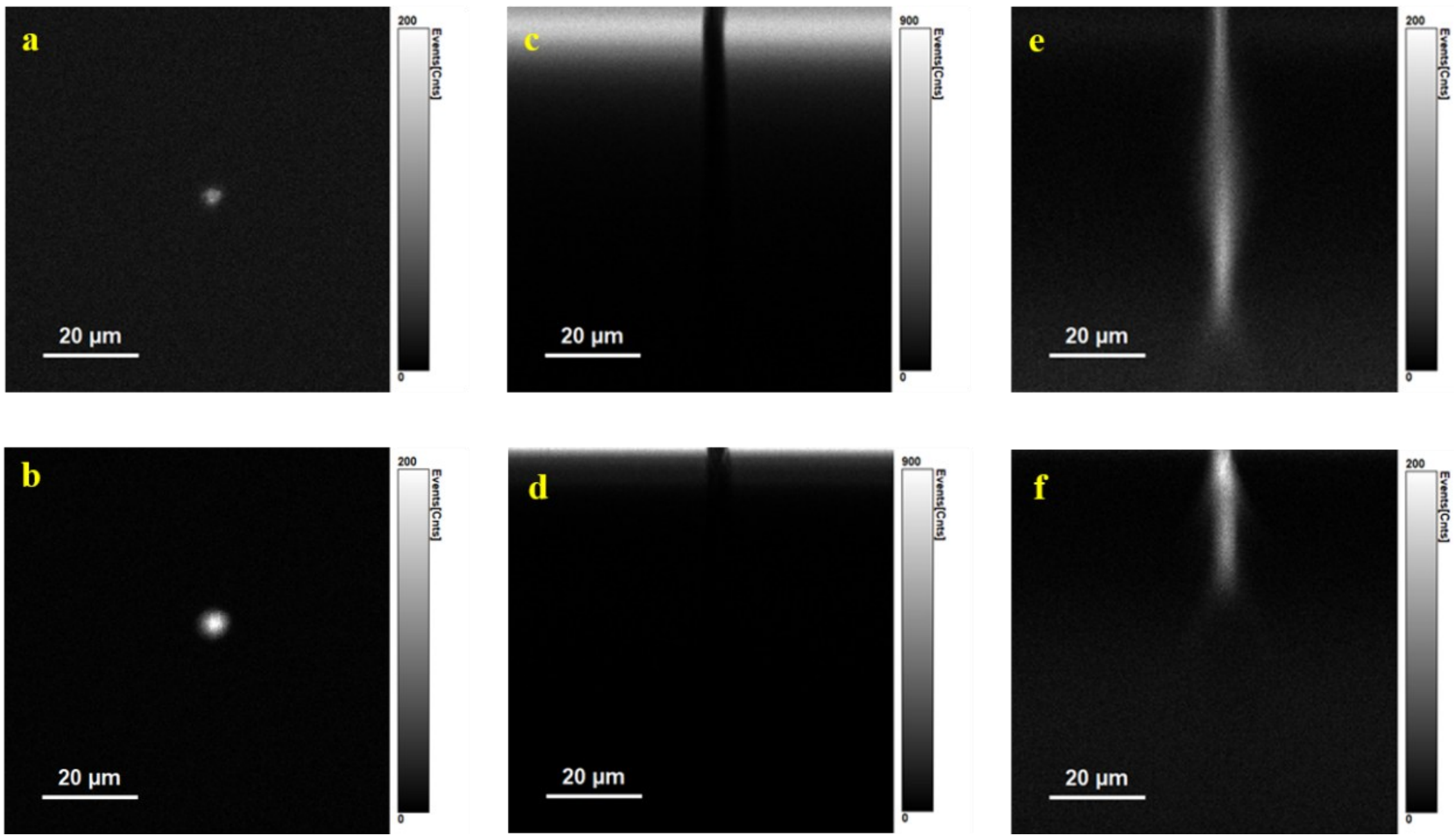

**Fig. 3| Micropipe sidewall mapping with different-NA objectives. (a), (b)** Lateral surface DPL maps of a micropipe acquired using the objectives with NA of 0.4 and 0.65, respectively. **(c)–(f)** Cross-sectional BS and DPL maps of the same micropipe, obtained by scanning the focal plane from ~5 μm above the surface into the bulk.

**DPL spectra of micropipe sidewall defects**

Our non-line-of-sight confocal multiple-reflection DPL technique provides direct, nondestructive optical access to the inner-sidewall defects in SiC micropipes, thereby providing a new opportunity for further probing their electronic properties and underlying defect physics. Fig. 4a presents the room-temperature DPL spectrum acquired from the micropipe sidewalls under an excitation power density of 690 W/cm$^2$ using an oil-immersion objective, which affords high collection efficiency while minimizing chromatic aberration. The DPL spectra are presented on an energy scale after Jacobian transformation, which is more appropriate for analysis of the defect physics (see Supplementary Figs. 1). The DPL spectrum exhibits a distinctly asymmetric and ultrabroad emission profile, characterized by a maximum near 2.05 eV, a weaker shoulder around 2.5 eV, and an extended high-energy tail. This unusually ultrabroad line-shape likely reflects the combined effects of homogeneous broadening associated with phonon coupling and inhomogeneous broadening arising from the high-density defect states that are broadly distributed in both space and energy.

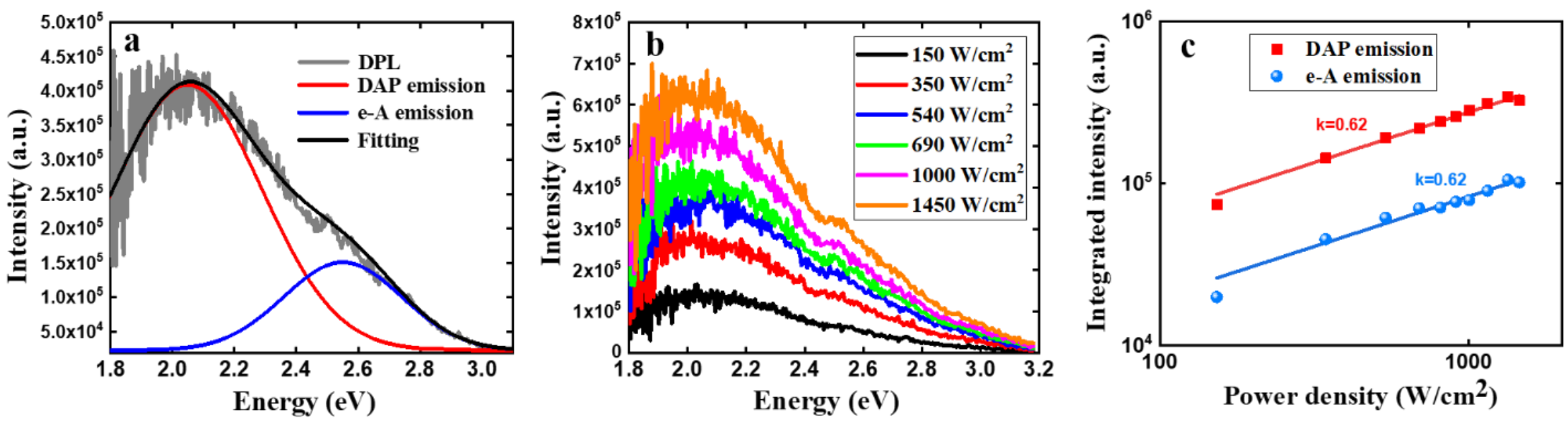

**Fig. 4| Electronic properties of micropipe sidewall defects. (a)** DPL spectrum of micropipe sidewall defects fitted with two Gaussian functions. (**b)** Power-dependent DPL spectra and **(c)** the corresponding plot of the DPL intensity versus the excitation power in log–log scale for both DAP and e-A emissions.

Defect-related luminescence in wide-bandgap semiconductors generally arises from two recombination channels: (i) free-to-bound recombination between conduction-band electrons and defect-localized holes (e–A transitions), and (ii) donor–acceptor-pair-like recombination between the donor-bound electrons and acceptor-bound holes [22-26]. The latter may originate either from externally introduced dopants or from donor-like and acceptor-like states associated with the same defect complex, such as the extended defects. We refer to the latter as inherent DAP-like emission to distinguish it from conventional dopant-induced DAP recombination (see Supplementary Figs. 2).

As a result, the ultrabroad DPL spectrum was deconvoluted into two Gaussian components, tentatively assigned to intrinsic DAP-like and e–A emissions, respectively (Fig. 4a). Notably, the DAP-like component remains dominant over the e–A component even at room temperature, in sharp contrast to conventional defect-related luminescence. In general, the DAP emission is favored at low temperatures, whereas at room temperature it is typically weak or absent, as it relies on carrier localization at both donor-like and acceptor-like states and is readily suppressed by thermally activated detrapping. This anomalous DAP-like emission will be further elucidated and confirmed through comprehensive analysis of power-dependent and time-resolved DPL measurements.

In both recombination processes, the participating deep-level states are strongly coupled with lattice vibrations, resulting in pronounced vibronic broadening and substantial spectral asymmetry. Furthermore, the extended geometry and structural inhomogeneity of the micropipe sidewalls give rise to an energetic distribution of defect states, as well as a spatial distribution of the DAP-state separations. These spatially and energetically distributed high-density defect states further broaden both the e–A and intrinsic DAP-like emissions, producing a superposition of recombination channels that yields the observed ultrabroad emission band at room temperature.

To further elucidate the underlying electronic transitions, we performed excitation-power-dependent DPL measurements. Fig. 4b presents representative DPL spectra for comparison. At the lowest excitation power density of 150 W/cm², the intrinsic DAP-like emission dominates the spectrum, accounting for approximately 82% of the total emission. Even at the highest excitation power density of 1450 W/cm², this component remains prevalent, contributing approximately 75%. Notably, the DAP-like emission exhibits a slight blue shift with increasing excitation power, which is a characteristic feature of DAP-type recombination. In contrast, the relative contribution of the e–A emission gradually increases with excitation power, while its emission energy remains nearly unchanged (see Supplementary Figs. 3). The integrated DPL intensity was further plotted as a function of excitation power on a log–log scale, revealing typical defect-related emission behavior with a power-law exponent of approximately 0.62 for both the intrinsic DAP-like and e–A emissions[27,28]. The similar power-law exponents suggest that the DAP-like and e–A emissions are governed by closely related recombination kinetics, likely involving a shared carrier reservoir.

**Carrier dynamics of two coupled emissions**

To further elucidate the defect physics of micropipe sidewalls, particularly the associated carrier dynamics, we performed time-resolved defect-PL measurements. Fig. 5a shows the decay profiles of the broadband defect-PL emission. Because the decay kinetics of the DAP emissions are strongly governed by the spatial separation and wavefunction overlap of the donor–acceptor pairs, the corresponding recombination rates can be written as, $\mathrm{W(r)} = W_{max}\exp(-\frac{2r}{a_0})$, where $W_{max}, r,$ and $a_0$ denote the maximum recombination rate, the donor–acceptor separation, and the effective Bohr radius, respectively. Accordingly, the transient decay profiles deviate markedly from single-exponential behavior due to the contributions of spatially and energetically distributed DAP states. To resolve the carrier dynamics of the individual recombination channels and clarify their interplay, a more detailed analysis based on spectrally selective TRPL measurements is required. We therefore performed spectrally resolved TRPL measurements for both the DAP-like and e–A emissions using appropriate bandpass filters, as shown in Fig. 5b.

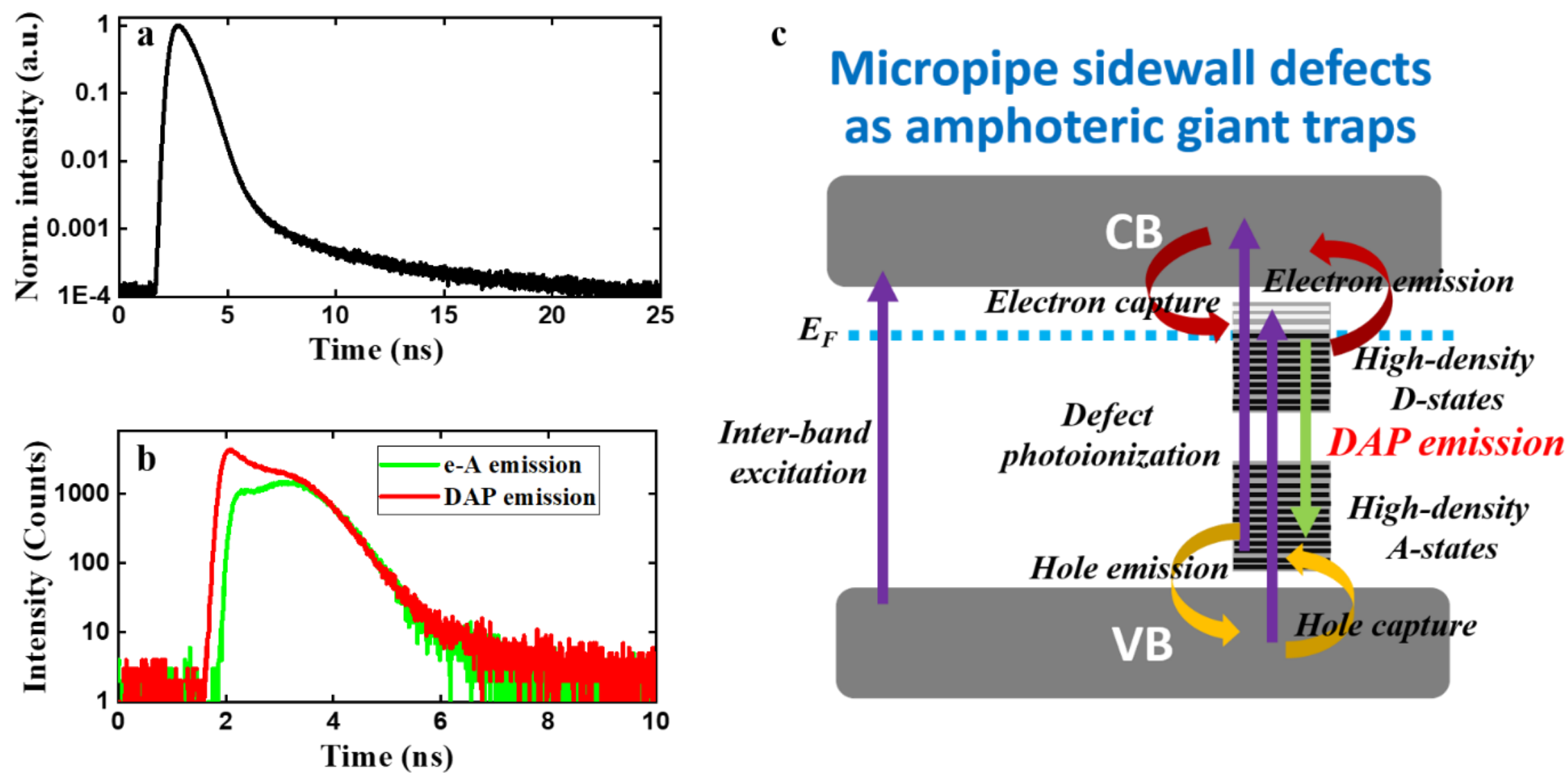

**Fig. 5| Carrier dynamics of coupled DAP and e–A emissions. (a)** Time-resolved DPL decay profiles of micropipe sidewall defects. **(b)** Individual decay profiles of the inherent DAP-like and e–A emissions, respectively. **(c)** Schematic energy band diagram of the micropipe sidewall defects, showing the intrinsic donor-like and acceptor-like states and the associated photoionization and recombination pathways.

Notably, the resulting transients exhibit distinctly different carrier dynamics for the two channels, while also revealing clear signatures of coupled behavior between them. The time-resolved DAP-like emission exhibits an unusual transient profile with rich underlying physics, consisting of a rapid rise, a fast nanosecond-scale decay, and an anomalously persistent plateau-like region followed by a slow tail. These features suggest that the inherent DAP states are populated rapidly and undergo efficient radiative recombination at early times, whereas the rebuilt-up plateau is consistent with sustained carrier refilling from a trapped-carrier reservoir associated with the high density of defect states on the micropipe sidewalls. In contrast, the e–A emission shows a pronounced delayed rise, reaching its maximum only after the DAP-like signal has begun to decay. In particular, the initial decay of the inherent DAP-like emission coincides with a concomitant rise of the e–A emission. This behavior suggests that the e–A channel is not populated directly upon photoionization, but rather through a secondary process, most likely involving detrapping-mediated carrier transfer from the donor-like sidewall defect states into the band-edge states.

At longer delay times, the DAP-like and e–A transients gradually converge and exhibit nearly identical decay behavior. This convergence suggests that the two emission channels are no longer governed by their individual intrinsic recombination rates, but instead by a common slow kinetic process. Such behavior is consistent with both channels being coupled to the same trapped-carrier reservoir associated with the dense defect manifold on the micropipe sidewalls. In this late-time regime, the observed decay is likely controlled by a quasi-steady-state balance among carrier trapping, detrapping, and recombination, rather than by independent channel-specific dynamics. This shared late-time decay is also consistent with the similar power-law exponents observed in the power-dependent DPL measurements, suggesting that both emission channels are governed by related carrier-supply and trapping–detrapping dynamics.

By combining the steady-state DPL spectra with the time-resolved decay dynamics, we can establish a physical model for the electronic properties of the micropipe sidewall defects in n-type SiC, as schematically illustrated in Fig. 5c. Under 375 nm laser excitation (~3.3 eV, near-band-edge excitation), both near-band-edge excitation of the SiC host and direct photoionization of defect states at the micropipe sidewalls can occur in our configuration. In conventional direct-bandgap semiconductors under above-bandgap excitation, photocarriers are primarily generated through inter-band transitions between the valence and conduction bands. In SiC, however, the indirect-bandgap nature of the host makes near-band-edge inter-band excitation optically less efficient under near-resonant excitation. Meanwhile, direct illumination of the defect-rich micropipe sidewalls can enhance defect-selective photoionization.

In n-type SiC epilayers, such defect photoionization likely proceeds via transitions from the valence band to empty donor-like states, and from occupied acceptor-like states into the conduction band (see Supplementary Figs. 4). The resulting non-equilibrium carriers are then rapidly captured by the donor- and acceptor-like sidewall states, leading to the fast formation of intrinsic DAP states and subsequent DAP-like emissions. This mechanism is consistent with the rapid rise times and nanosecond-scale decay observed in our TRPL measurements. These spectroscopic signatures, together with the observed coupled carrier dynamics, suggest that micropipe sidewalls can act as amphoteric giant traps associated with extended defects, facilitated by their dense manifolds of both donor- and acceptor-like states.

Recently, Chen et al. employed ultrabroadband transient absorption spectroscopy to investigate carrier capture and recombination dynamics of carbon-related defects in GaN[29]. They found that negatively charged $C_N$ point defects possess a large hole-capture coefficient, resulting in a very short recombination lifetime. Similarly, Jiang et al. theoretically investigated the carrier-capture pathways of carbon vacancies in 4H-SiC using first-principles calculations[30]. Their work showed that $Z_{1/2}$ centers with negative-U behavior exhibit large electron and hole capture cross-sections, thereby acting as highly efficient lifetime killers. However, giant-trap behavior associated with extended defects remains largely unexplored experimentally. Owing to their large effective carrier-interaction area and dense defect-state manifolds, amphoteric giant traps associated with extended defects can play a critical role in device operation. Our identification of intrinsic DAP-like emission from micropipe sidewalls provides direct experimental evidence for such giant-trap behavior in extended defects, which has not been clearly demonstrated previously.

## DISCUSSION

### Micropipe sidewall defects as an alternative leakage mechanism

Despite the consensus that micropipes are "killer defects" in SiC power devices, their leakage mechanism remains incompletely understood and is often attributed solely to geometric electric-field crowding. However, field crowding alone is insufficient to explain their electrical behavior. If it were the dominant and universal mechanism, other high-curvature microstructures, such as voids and pits, would be expected to exhibit comparable leakage currents. Furthermore, local field enhancement cannot readily account for the significant leakage currents observed under low reverse bias, nor the relatively benign electrical behavior of most open-core threading screw dislocations (nanopipes), which are dissociation products of micropipes[31,32]. This discrepancy suggests that the catastrophic failure associated with micropipes is governed not only by their geometry, but also by the defect physics of their internal sidewall surfaces.

The observed carrier dynamics and recombination kinetics establish a direct link between the defect physics of micropipe sidewalls and the macroscopic leakage behavior of the device. These micropipe-associated amphoteric giant traps, characterized by a dense manifold of defect states, can efficiently capture and transiently store the charge carriers. Under reverse bias, these spatially distributed states facilitate field-enhanced trap-assisted conduction by the mechanisms such as trap-assisted tunneling and Poole–Frenkel emission. By providing a preferential leakage pathway along the internal surfaces of micropipe sidewalls, these states can lead to premature breakdown and the large leakage currents observed experimentally.

Although micropipe density has been substantially reduced in state-of-the-art SiC, these defects persist, particularly in the large-diameter, thick epitaxial wafers required for high-voltage power devices. Even a single micropipe can induce catastrophic breakdown. While most previous studies have focused on the formation mechanisms and structural characteristics of micropipes, the microscopic defect nature of their inner sidewalls and the underlying defect physics remain largely unexplored. Recently, Lee et al. investigated the micropipes in 200 mm 4H-SiC epitaxial wafers using selected-area electron diffraction and energy-dispersive X-ray spectroscopy, revealing Si-rich regions along the inner sidewalls[6]. These findings suggest that micropipe sidewalls constitute a chemically and electronically distinct environment, further highlighting the unique material properties of these defective regions.

### Activation of deep-level defect emissions

It is worth noting that deep-level defects are commonly regarded as efficient non-radiative recombination centers. These centers often trap charge carriers and relax via multiphonon emission or function as Shockley–Read–Hall centers, sequentially capturing the electrons and holes to facilitate recombination without photon emission. Furthermore, local strain can enhance electron–phonon coupling at these sites, further strengthening non-radiative relaxation pathways. Consequently, deep-level defects are generally inefficient light emitters and are frequently characterized as "PL-dark" centers, which is typically inferred only indirectly through the quenching of near-band-edge emission. As a result, direct electronic information regarding the nature of these defects remains remarkably difficult to access.

Defect-related emission from the deep-level states typically involves carrier generation within the host bands followed by capture into defect states. The efficiency of this process is governed by defect-state density and specific capture coefficients, which must compete with direct band-to-band recombination, thermal detrapping, and other defect pathways. To effectively activate the deep-level emission from specific centers, we employ direct defect photoionization. This approach generates carriers with lower excess kinetic energy, thereby suppressing hot-carrier-assisted relaxation and providing a more selective probe of defect-mediated dynamics by minimizing carrier thermalization. Furthermore, unlike conventional impurity-induced DAPs, the donor- and acceptor-like defect states in this system are co-located within the same defect complex (i.e. micropipe sidewalls). Their extreme spatial proximity and enhanced wavefunction overlap are expected to promote efficient, inherent DAP-like recombination.

In conclusion, we have developed a non-line-of-sight confocal multiple-internal-reflection spectromicroscopy to directly probe the electronic properties of micropipe sidewalls. Through comprehensive analysis of steady-state and time-resolved defect-PL behavior, we reveal that the sidewall defects exhibit ultra-broad emission bands arising from both inherent DAP-like and e-A transitions. The inherent DAP-like emission was characterized by a rapid rise, nanosecond-scale decay, and a unique rebuilt plateau in the carrier dynamics. Our findings suggest that the micropipe sidewall defects can serve as extended amphoteric giant traps and charge reservoirs, thus facilitating leakage current transport via trap-assisted tunneling and Poole–Frenkel emission mechanisms. Our work provides a non-destructive optical technique for directly accessing the electronic nature of high-aspect-ratio hollow-type defects, while unraveling the fundamental physics that can trigger catastrophic leakage in SiC power devices.

## METHODS

### Sample Preparation

The samples were obtained from a commercial 6-inch n-type 4H-SiC epitaxial wafer (BEST Compound Co., Ltd.; SICC substrate). The epitaxial layers were grown by chemical vapor deposition along the [0001] direction on a 4° off-axis substrate. The nitrogen-doped substrate exhibited a resistivity of $\rho$=0.015–0.028 Ω·cm and a thickness of 350 μm. The epitaxial structure consisted of a heavily doped buffer layer (~2 μm, Nd = $9.3 \times 10^{17}$ $cm^{-3}$), a lightly doped drift layer (~60 μm, $1.2\times10^{15}$ $cm^{-3}$), and a current-spreading layer (~1 μm, $2.38\times10^{16}$ $cm^{-3}$). The epi-surface roughness was $R_a$ = 0.14 nm. For optical measurements, the wafer was cleaved into ~1 × 2 $cm^2$ specimens along the primary flat direction to facilitate handling on the confocal piezoelectric stage. Each specimen was subjected to standard RCA-1 cleaning immediately before measurement.

**Non-line-of-sight confocal multiple-reflection defect-PL Spectromicroscopy**

Non-line-of-sight confocal multiple-reflection defect photoluminescence (DPL) spectromicroscopy was implemented on a laser-scanning confocal microscope (MT-100, PicoQuant, Germany) coupled to a spectrophotometer. Excitation was provided by a 375 nm picosecond pulsed diode laser (LDH-P-C-375, PicoQuant) and delivered through a laser combining unit (LCU) with neutral-density attenuation and single-mode fiber coupling. The incident excitation power at the sample plane was monitored using a power meter equipped with a calibrated photodiode sensor and was varied via the attenuator for excitation-power-dependent measurements.

PL detection was operated in two complementary modes. For confocal imaging, two single-photon avalanche diodes (SPADs) were used in parallel with a 50/50 beam splitter: one equipped with a 377 nm bandpass filter for laser backscattering detection, and the other with a 400 nm long-pass filter for defect-related PL, enabling simultaneous BS and PL mapping. For spectral measurements, the emission was fiber-coupled into the spectrophotometer and detected using a thermoelectrically cooled hybrid photomultiplier tube. Three objectives mounted on a three-axis piezoelectric stage were employed. Dry Olympus PlanN 20×/0.40 and PlanN 40×/0.65 plan achromats were used for surface, subsurface, and cross-sectional confocal mapping of micropipes. High-resolution spectral and time-resolved measurements at fixed focal positions were performed using an Olympus UPlanSApo 100×/1.40 oil-immersion super-apochromat (Olympus Corporation, Japan). The super-apochromatic correction (~400–700 nm) ensures a consistent focal volume across both DAP-like and e–A emission bands.

For time-resolved PL measurements, time-correlated single-photon counting (TCSPC) electronics were employed. The instrument response function (IRF), measured by collecting laser backscattering light, was approximately 200 ps, enabling reliable time resolution of nanosecond-scale carrier dynamics. Spectrally selective TRPL measurements were performed using appropriate bandpass filters to isolate the DAP-like and e–A emission channels.

## DATA AVAILABILITY STATEMENT

The data that support the findings of this study are available from the corresponding author upon reasonable request.

## ASSOCIATED CONTENT

### Supplementary Information

Supplementary Information (S1–S4) is available right below the references.

## AUTHOR INFORMATION

**Corresponding Author**

*E-mail：wenchunglee@waferworks.com (Wen-Chung Li); ctyuan@cycu.edu.tw (Chi-Tsu Yuan)

**ORCID**

Chi-Tsu Yuan：[0000-0003-3790-9376](0000-0003-3790-9376)

**AUTHOR CONTRIBUTIONS**

Chi-Tsu Yuan and Wen-Chung Li supervised the project and led the research direction. They conceived the idea and designed the experiments. Irwan Saleh Kurniawan and Russel Cruz Sevilla designed the techniques, performed the experiments, and carried out the data analysis. Ruth Jeane Soebroto and Hsiu-Ying Huang contributed to in-depth discussions. Ji-Lin Shen and Sheng Hsiung Chang provided valuable suggestions and assisted with manuscript preparation. Chi-Tsu Yuan wrote the manuscript. All authors discussed the results, contributed to the manuscript, and approved the final version for submission.

**NOTES**

The authors declare no competing financial interest.

**ACKNOWLEDGEMENTS**

This work was supported by the National Science and Technology Council under the grant number NSTC 113-2112-M-033-007-MY3 and 113-2112-M-033-003.

**Supplementary Information**

# Unraveling the Defect Physics of SiC Micropipe Sidewalls by Non-Line-of-Sight Confocal Spectromicroscopy: Amphoteric Giant Traps

Irwan Saleh Kurniawan[#1,2], Russel Cruz Sevilla[#1,2], Ruth Jeane Soebroto[1,2], Hsiu-Ying Huang[1,2], Hsiu-Ming Hsu[1,2], Ji-Lin Shen[1,2], Sheng Hsiung Chang[3], Wen-Chung Li*[1,2,4] Chi-Tsu Yuan*[1,2]

[1]Department of Physics, Chung Yuan Christian University, Taoyuan, Taiwan

[2]Research Center for Semiconductor Materials and Advanced Optics, Chung Yuan Christian University, Taoyuan, Taiwan

[3]Department of Optics and Photonics, National Central University, Taoyuan, Taiwan

[4]WAFER WORKS, Taoyuan, Taiwan

## SI 1. DPL spectra represented by the energy scale by Jacobian transformation

For PL measurements aimed at defect analysis, presenting PL spectra on an energy scale is generally more appropriate than using a wavelength scale, because defect-related optical transitions are fundamentally governed by their electronic transition energies. In semiconductor defect physics, key physical parameters such as bandgap energies, defect levels, donor–acceptor pair transition energies, phonon coupling, and activation energies are all naturally described in units of electron volts (eV). Moreover, plotting DPL spectra on an energy scale can preserve the intrinsic spectral line-shape, which otherwise would be distorted on a wavelength scale owing to the nonlinear relationship between energy and wavelength [1]. This consideration is particularly important for broad defect-related emission bands, as encountered in the present case.

Accordingly, presenting PL spectra as a function of photon energy not only enables a more direct interpretation of the underlying electronic structure and defect-related recombination processes, but also provides a more reliable representation of the intrinsic DPL line-shape. For this purpose, the raw wavelength-domain spectra were converted into energy-domain spectra using the Jacobian transformation associated with the nonlinear relation, $\mathrm{f(E)} = \mathrm{f(\lambda)}\frac{d\lambda}{dE} = -f(\lambda)\frac{hc}{E^2}$. This transformation corrects the distortion that would otherwise arise from the nonlinear wavelength-to-energy conversion and thus preserves the intrinsic-line shape of the broad defect-related emission bands.

## SI 2. Inherent DAP-like emissions

Unlike conventional DAP emissions associated with external impurities, inherent DAP-like emissions originate from radiative recombination between defect-related donor states and defect-related acceptor states. Such donor- and acceptor-like defect states may exist in both bulk semiconductors and low-dimensional quantum dots (QDs) [2–4]. For example, colloidal AgInS2 QDs show dominant defect-related DAP-like emission under above-bandgap excitation. In this case, band-edge emission is absent because both photogenerated electrons and holes are efficiently captured by the defect states before band-edge recombination occurs. The recombination lifetime of such defect-related DAP-like emissions in nanoscale QDs is typically on the microsecond timescale, depending on the defect concentration.

**SI 3. DPL spectra under low and high excitation power**

As the excitation power increases, the relative contributions of the intrinsic DAP-like and e–A emissions change systematically. To quantify this behavior, the DPL spectra measured at the lowest and highest excitation powers were fitted using two Gaussian components, as shown below. At the lowest excitation power, the DAP-like emission dominates, accounting for 82% of the total emission. Even at the highest excitation power, it remains the dominant component, contributing up to 76%. With low excitation power, photogenerated carriers are efficiently captured by micropipe sidewall defects, thereby favoring DAP-like recombination. As the excitation power increases, the DAP-like emission remains dominant, likely owing to the high density of donor- and acceptor-like defect states on the micropipe sidewalls.

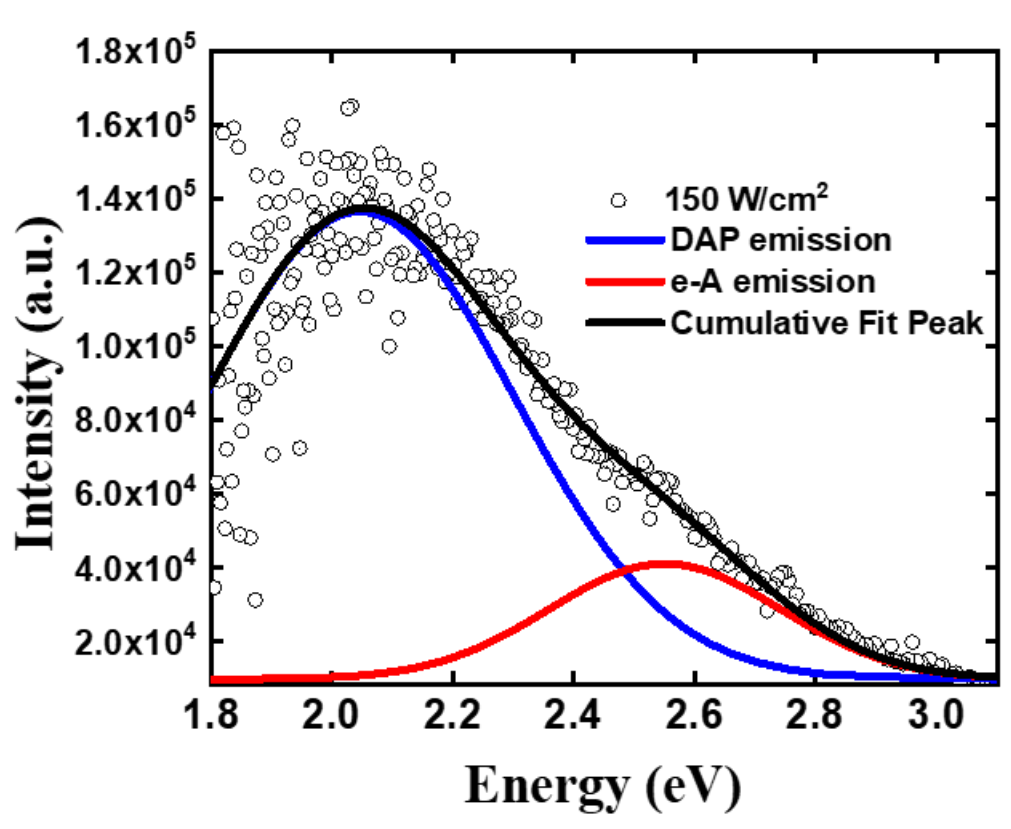

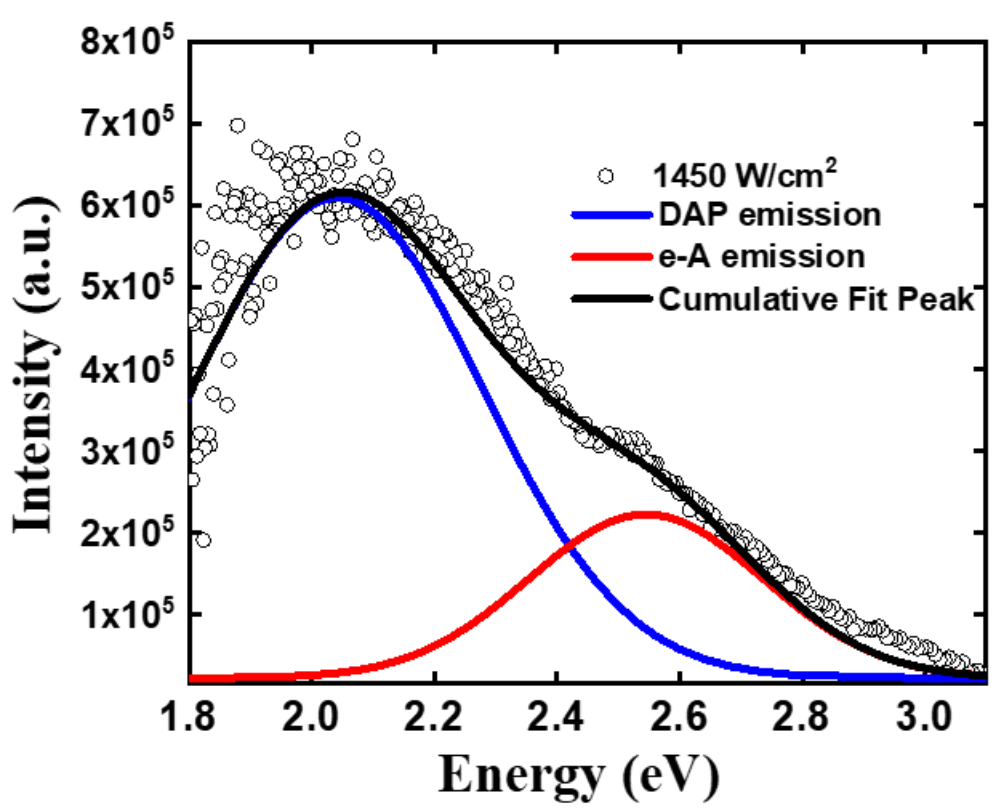

**S4. Direct photoionization of micropipe sidewall defects**

Due to the indirect-bandgap nature of 4H-SiC, excitation at 375 nm gives rise to only weak phonon-assisted band-to-band absorption in the bulk. Moreover, under our confocal excitation geometry, the laser beam is deliberately focused into the hollow core of the micropipe, so that it directly overlaps with the defects located on the inner sidewalls rather than being predominantly absorbed near the planar SiC surface. Because the inner sidewalls host a high density of deep-level defect states, defect-related photoionization can become a significant, and likely dominant, excitation pathway under these conditions at this wavelength. If the band-to-band excitation were the dominant process, the defect-related emissions would be expected to occur only after photogenerated carriers diffuse and are captured by the nearby defect states, which is generally a much slower process. In many cases, such carrier capture occurs on ~microsecond or even longer timescales, depending on the defect nature and environment [5, 6]. By contrast, our time-resolved DPL measurements of the DAP emissions exhibited a prompt rise, which strongly suggests that direct photoionization of the sidewall defects plays a dominant role in the excitation process.